\documentclass[preprint,showpacs,preprintnumbers,amsmath,amssymb,showkeys]{revtex4}

\usepackage{graphicx}
\usepackage{dcolumn}
\usepackage{bm}

\newcommand{\ddt}{\frac{d}{dt}}
\newcommand{\ds}[1]{\displaystyle{#1}}
\newcommand{\uk}{{\hat \vectu}_\vectk}

\newcommand{\fk}{{\hat \vectf}_\vectk}

\newcommand{\psik}{{\hat \vectpsi}_\vectk}
\newcommand{\lk}{\ell |\vectk|}
\newcommand{\sumk}{\sum_\vectk}
\newcommand{\ave}[1]{\left\langle {#1} \right\rangle}
\newcommand{\dissipation}{\beta}

\newcommand{\vectzero}{{\mathbf{0}}}

\newcommand{\vectf}{{\mathbf{f}}}

\newcommand{\vectk}{{\mathbf{k}}}

\newcommand{\vectu}{{\mathbf{u}}}

\newcommand{\vectx}{{\mathbf{x}}}

\newcommand{\vectphi}{{\boldsymbol{\phi}}}

\newcommand{\vectpsi}{{\boldsymbol{\psi}}}

\newcommand{\rmd}{{\mathrm{d}}}
\newcommand{\rme}{{\mathrm{e}}}

\newcommand{\rmi}{{\mathrm{i}}}

\newcommand{\bequ}[1]{\begin{equation}\label{#1}}
\newcommand{\eequ}{\end{equation}}
\newcommand{\barr}[1]{\begin{eqnarray}\label{#1}}
\newcommand{\earr}{\end{eqnarray}}
\newcommand{\barrz}{\begin{eqnarray*}}
\newcommand{\earrz}{\end{eqnarray*}}

\newcommand{\NPnote}[1]{}

\begin{document}

\preprint{APS/123-QED}

\title{Energy Dissipation in Fractal-Forced Flow}

\author{Alexey Cheskidov}
 \email{acheskid@umich.edu}
\affiliation{%
Department of Mathematics, University of Michigan, Ann Arbor, MI 48109 
}%

\author{Charles R.\ Doering}%
 \email{doering@umich.edu}
\affiliation{
Department of Mathematics  \& Michigan Center for Theoretical Physics\\
University of Michigan, Ann Arbor, MI 48109 
}%

\author{Nikola P.\ Petrov}%
 \email{npetrov@ou.edu}
\affiliation{
Department of Mathematics
University of Oklahoma, Norman, OK 73019
}%

\date{\today}

\begin{abstract}
The rate of energy dissipation in solutions of the body-forced $3$-$d$ incompressible
Navier-Stokes equations is rigorously estimated with a focus on its dependence
on the nature of the driving force.
For square integrable body forces the high Reynolds number (low viscosity)
upper bound on the dissipation is independent of the viscosity, consistent with 
the existence of a conventional turbulent energy cascade.
On the other hand when the body force is not square integrable, i.e., when 
the Fourier spectrum of the force decays sufficiently slowly at high wavenumbers, there is 
significant direct driving at a broad range of spatial scales.
Then the upper limit for the dissipation rate may diverge at high Reynolds numbers,
consistent with recent experimental and computational studies of ``fractal-forced'' turbulence.
\end{abstract}

\maketitle


\section{Introduction}

A fundamental principle of modern hydrodynamic turbulence  theory is that 
nonlinear interactions between Fourier modes of the velocity field can transfer 
energy from directly-forced large spatial scales, through the so-called inertial 
range, down to a small dissipation length scale where viscosity effectively
consumes kinetic energy and transforms it into heat.  
This turbulent cascade process has been intensively
studied experimentally, numerically, and theoretically (at various levels of
mathematical rigor) since the first half of the twentieth century.
See, e.g., the book by Frisch \cite{Frisch95} for an introduction and entry into
the vast literature on this subject, which is still the focus of much current research.  

One profound consequence of the cascade mechanism is the so-called dissipative anomaly
wherein a finite  and non-vanishing residual energy dissipation persists in the singular limit
of vanishing viscosity, i.e., in the infinite Reynolds number limit.
This phenomenon is quantitatively described as Kolmogorov scaling of
the energy dissipation, namely 
\begin{equation} 
\dissipation \  \equiv \  \frac{\epsilon \ell}{U^3} \ = \ {\cal O}({Re^{0}}) \quad \text{as } Re \rightarrow \infty
\end{equation}
where $\epsilon$ is the total energy dissipation rate per unit mass, $\ell$ is an integral (large)
length scale in the flow characterizing the domain or a large scale in the forcing and flow, 
$U$ is a turbulent velocity scale, and $Re = U\ell/\nu$ is the Reynolds
number with $\nu$ denoting the kinematic viscosity.
Sreenivasan has collected together relevant data illustrating
Kolmogorov scaling in experiments \cite{Sreenivasan84}
and direct numerical simulations \cite{Sreenivasan98}.
Moreover, given precise definitions of all the quantities involved, this $\dissipation \sim Re^{0}$ 
Kolmogorov scaling has been shown to be an upper bound for (weak) solutions of the
$3$-$d$ incompressible Navier-Stokes equations driven by sufficiently smooth---in
particular, square integrable---body forces \cite{DoeringF02,DoeringES03,PetrovLD05,DoeringP05}.

While the cascade picture of turbulence requires that energy be predominantly injected 
in a relatively narrow range of spatial scales, some researchers have recently 
performed experimental and computational studies of 
{\em fractal-generated turbulence}.
These are flows 
driven by spatially broadband {\em fractal forces} with certain scaling properites
that inject energy directly at a wide range of scales---most notably at small
scales that could otherwise only be excited by the cascade.
Such forcing can impose a self-similar structure on the flow that is
independent of the turbulent energy cascade.  
If such forcing can be achieved experimentally then one can observe, and in
principle control, the balance between the energy that has been
directly injected and the energy transfered by the
nonlinear mode interactions.  

Indeed, Queiros-Conde and Vassilicos \cite{Queiros-CondeV01} performed experiments by
forcing fluid past a fractal object, an obstacle that was structurally self-similar over several scales.
Staicu {\it et al} \cite{StaicuMVW03} experimentally measured energy spectra and
structure functions in the wake of grids of fractal dimensions 2.05, 2.17, and 2.40 in
a wind tunnel, concluding that ``there may be a direct relation between the
scaling properties of the fractal object and the turbulence it creates''.  
This is more easily investigated in direct numerical simulations where details of the flow field is directly observable.

Mazzi and Vassilicos \cite{MazziV04} performed direct numerical
simulations of stationary homogeneous and isotropic turbulence 
in a fluid in a $3$-$d$ periodic box of size $\ell$ driven 
by a velocity-dependent fractal body force $\vectf(\vectx,t)$ with Fourier components 
of the form: 
\begin{equation}  \label{eq:Fk-MV}
\fk(t) 
=
\left\{
\begin{array}{ll}
F \, (\lk)^\zeta \ds{\left(\frac{\uk(t)}{|\uk(t)|} 
+ \rmi \, \frac{\vectk}{|\vectk|}\times\frac{\uk(t)}{|\uk(t)|} 
	\right)} \ ,
		&0 < |\vectk| < k_F \ , \\[2mm]
0 \ ,& 	|\vectk| > k_F 
\end{array}
\right.
\end{equation}
where $\uk$ are the velocity field's Fourier components, $\fk=\vectzero$ whenever $\uk=\vectzero$,
and ${\hat \vectu}_\vectzero \equiv \vectzero$.
The scaling exponent $\zeta$ is intended to characterize the fractal properties 
of the stirrer or obstacle, and the maximum wavenumber $k_F$ 
is to be thought of as the inverse of the spatial size
of the smallest parts of the fractal stirrer.  
Mazzi and Vassilicos used numerical values for which 
the fractal forcing extended down to scales $\sim k_F^{-1}$ on the
order of the Kolmogorov dissipation length $\eta \equiv (\nu^3/\epsilon)^{1/4}$.
They observed that the bulk energy dissipation rate did not
exhibit Kolmogorov scaling $\dissipation \sim Re^0$, but rather
$\dissipation \sim Re^1$ corresponding to $\epsilon \sim U^4/\nu$.

Biferale {\it et al} \cite{BiferaleLT04,BiferaleCLST04} performed numerical simulations
of the $3$-$d$ Navier-Stokes equations with a stochastic body force that was 
white-noise in time but with a power law spectrum of spatial scales $\sim k^\zeta$.
They investigated small scale turbulent fluctuations and concluded that the statistics
displayed two distinct qualitative behaviors.
When the spatial spectrum of the forcing decayed sufficiently fast, 
the small scale fluctuations were universal in the sense that they were 
independent of the details  of the force spectrum.
This regime corresponds to conventional cascade dynamics.
When the spatial spectrum of the forcing decayed more 
slowly, however, the small scale fluctuations were ``force-dominated''
with the cascade being overwhelmed by the
direct excitation from the driving.
Interestingly, they reported that this transition occurs at a value $\zeta = -\frac32$
of the scaling exponent corresponding to the boundary between (spatially) 
square integrable and ``rougher'' forcing functions without square
summable Fourier coefficients.

In this paper we derive rigorous upper bounds on the bulk energy dissipation 
$\epsilon$ in
an incompressible Newtonian fluid driven by a variety of body forces
including forces that are {\em not} square integrable.
This work generalizes the previous analysis for square integrable body forces 
\cite{DoeringF02,DoeringES03,DoeringP05} to include fractal forces
that drive the flow directly at a broad range of scales.
In accord with the findings of Biferale {\it et al} we find that the case of
square integrable forcing is a borderline situation: $\dissipation \lesssim Re^{0}$ 
when the body forces are square integrable (or smoother), but the estimates
increase for rougher driving so that the dissipation coefficient $\dissipation$ may
increase asymptotically as $Re \rightarrow \infty$.
For the roughest forcing functions that make sense mathematically, i.e., forcing functions
with Fourier coefficients satisfying $\sum_\vectk k^{-2} |\fk|^{2} < \infty$, we find that
$\dissipation \lesssim Re^{1}$, the scaling observed by Mazzi {\it et al} .

The rest of this paper is organized as follows.
The following Section II lays out the mathematical setting for the analysis and gives the
definitions of the physically relevant quantities of interest.
In Section III we study the case of time-independent body forces,
and the subsequent Section IV deals with velocity-dependent forces like 
\eqref{eq:Fk-MV}.
The concluding Section V contains a brief summary and some closing remarks.
For completeness and to make the paper self-contained,
we include some mathematical details in an appendix.

\section{Statement of the problem and definitions}

Consider the incompressible $3$-$d$ Navier-Stokes equations on a periodic domain
$\vectx \in [0,\ell]^3$:
\begin{equation} \label{NSE}
\partial_t \vectu + (\vectu \cdot \nabla)\vectu + \nabla p =  \nu \Delta \vectu + \vectf\\
\end{equation}
where $\vectu(\vectx,t)$ is the divergence-free velocity field,
$p(\vectx,t)$ is the pressure, $\vectf(\vectx,t)$ is
the applied body-force, $\nu>0$ is the kinematic viscosity, and
$\vectu|_{t=0}=\vectu_0(\vectx)$ is the initial condition.
We will take the body force to be a specified time independent (divergence-free) function
$\vectf(\vectx)$ in Section III, or given by a velocity-dependent expression like
\eqref{eq:Fk-MV} in Section IV. 
We write Fourier expansions as
\begin{equation}
\vectu(\vectx,t) = \sum_{\vectk} \uk(t) \rme^{i \vectx \cdot \vectk}
\quad \text{where} \quad 
\uk(t) = \frac{1}{\ell^3} \int_{\ell^3} \rme^{- i \vectx \cdot \vectk} \vectu(\vectx,t) d^3x.
\end{equation}
Without loss of generality, in every case we will take the applied body force and
initial data to have spatial mean zero so that $k = |\vectk| > 0$ in all sums.

A field $\vectu(\vectx,t) \in H^\alpha$ if $\|\vectu(\cdot,t)\|_{H^\alpha}<\infty$ 
where we define the Sobolev norms $\|\cdot\|_{H^\alpha}$ by 
\begin{equation}  \label{eq:Sob-standard}
\|\vectu(\cdot,t)\|_{H^\alpha}^2  \ \equiv \  \sum_\vectu (\ell k)^{2\alpha} |\uk(t)|^2
\ = \ \frac{1}{\ell^3} \int_{\ell^3} |(-\ell^2 \Delta)^{\alpha} \vectu(\vectx,t)|^2 d^3x.
\end{equation}
The index $\alpha$ can be positive or negative and the function spaces $H^\alpha$ are
nested according to $H^\alpha \subset H^{\alpha'}$ for $\alpha > \alpha'$.
The case $\alpha = 0$ corresponds to the usual $L^2$ norm
(with the volume normalization) and we write
\begin{equation}
\|\vectu\|_{H^0} = \|\vectu\|_{L^2} = \|\vectu\|
\end{equation}

``Fractal" forces are defined as those with power-law Fourier coefficients, $|\fk| = C k^\zeta$.
For such a function to belong to the Sobelov space $H^\alpha$ its exponent must
satistfy $\zeta<-\alpha - \frac32$, i.e., the Fourier coefficients must decay as
$|\fk| \lesssim k^{-\alpha - \frac32 - \delta}$ for some $\delta > 0$.

We define time averages of functions $g(t)$ according to
\begin{equation}
\overline{g} = \lim_{T \rightarrow \infty} \int_0^T g(t) dt
\end{equation}
and for simplicity in this paper we presume that this limit exists for all quantities of interest.
The bulk (volume and time) average of a function $h(\vectx,t)$ is denoted by
\begin{equation}
\ave{h} = \overline{\frac{1}{\ell^3} \int_{\ell^3} h(\vectx,\cdot) d^3x}.
\end{equation}

The root means square velocity $U$ of a solution $\vectu(\vectx,t)$
of the Navier-Stokes equations is
\begin{equation}
U = \ave{|\vectu|^2}^{1/2} = \overline{\| \vectu \|^2 }^{1/2},
\end{equation}
and the bulk energy dissipation rate (per unit mass) is defined by
\begin{equation}
\epsilon = \ave{ \nu |\nabla \vectu|^2} =  \overline{ \frac{\nu}{\ell^2} \| \vectu \|^2_{H^1}}.
\end{equation}
When a solution $\vectu(\vectx,t)$ satisfies the energy equality (i.e., when the energy
is absolutely continuous, which holds for every regular solution), 
the energy dissipation rate satisfies
\begin{equation}
\epsilon = \ave{\vectf \cdot \vectu}.
\label{PB}
\end{equation}
That is, the power supplied by the driving force is balanced by the viscous dissipation.

Weak solutions to these $3$-$d$ Navier-Stokes equations 
exist for $\vectf \in H^{-1}$, and then in general 
the relation in (\ref{PB}) is only an inequality,
i.e., $\epsilon \le \ave{\vectf \cdot \vectu}$
\cite{CFbook,DoeringG95,FoiasMRT01-book}.
This fact does not affect our results, however, because
we will just derive upper limits on $\epsilon$.
Moreover, the assumption of the existence of the 
long times averages is not necessary if the 
limit is replaced by $\limsup_{T\rightarrow\infty}$.
With that definition the estimates we derive are fully applicable to weak solutions.

Using the definitions above, the Reynolds number is identified $Re = U\ell/\nu$
and the dissipation coefficient as $\dissipation = \epsilon \ell / U^3$.
In the scenario described here both $Re$ (or $U$) and $\dissipation$ (or $\epsilon$)
are formally ``emergent'' quantities, not directly controllable but determined rather 
as functions of $\ell$ and $\nu$ and functionals of $\vectu_0$ and $\vectf$.
These bulk averaged quantities generally depend on $\vectu_0$, 
but the relationships derived below are uniformly valid for all  solutions
regardless of initial data so we will drop any further reference to them.
In practice one assumes that the parameters of the force, e.g., its amplitude,
can be tuned to achieve any desired Reynolds number.
Then $\dissipation$ may be considered a function of $Re$.
The overdamped highly viscous limit is $Re \rightarrow 0$ and the
 vanishing viscosity limit is explored as 
$Re \rightarrow \infty$.

Some very general statements can be made for the overdamped limit.
Poincare's inequality implies that
\begin{equation}
\epsilon \ge \frac{4 \pi^2 \nu}{\ell^2} U^2,
\end{equation}
so for any forcing
\begin{equation}
\beta \ge \frac{4 \pi^2}{Re}.
\label{lower}
\end{equation}
This Reynolds number scaling is sharp: as will be seen below, for a wide variety of
forces there exists a constant $c \ge 4\pi^2$ (generally depending
on the details of the forcing) such that 
\begin{equation}
\beta \le \frac{c}{Re} \quad \text{as} \quad Re \rightarrow 0.
\end{equation}
This scaling, $\beta \sim Re^{-1}$, is characteristic of large scale laminar flows
where the typical rate of strain is proportional to $U/\ell$
and the typical stress is proportional to $\nu U/ \ell$.

For higher Reynolds numbers the lower estimate in (\ref{lower}) can generally 
not be improved.
That is, at arbitrarily high $Re$ there are forces that can sustain the
flow with $\beta \sim Re^{-1}$.
Those flows---which may be unstable---are necessarily sufficiently laminar to exclude any
characterization as being turbulent.
The upper bound on $\dissipation$, however, necessarily increases above
$Re^{-1}$ as $Re \rightarrow \infty$.
For turbulent flows with an effective energy cascade the dissipation becomes independent of 
$\nu$ as $Re \rightarrow \infty$, i.e., $\beta \sim Re^0$, as evidenced by
experiments and direct numerical simulations.
But for sufficiently broadband forcing $\beta$ may increase indefinitely in this limit, and
the task of the next two sections is to place rigorous upper bounds on $\beta$
as a function of $Re$ for flows driven by fractal forces.


\section{Steady $H^{-\alpha}$ body forces}

In this section we generalize the approach introduced by Doering \& Foias 
\cite{DoeringF02}---an approach that was inspired by previous work of
Foias and coworkers \cite{FoiasMT93,Foias97,FoiasMRT01b}---to
cases where the time independent force $\vectf(\vectx)
\in H^{-\alpha}$ with $\alpha \in [0,1]$.
For $\alpha \le 0$ the force $\vectf \in L^2$ and the $\beta \lesssim Re^0$
upper bound, corresponding to the usual energy cascade, 
is effective \cite{DoeringF02}.
We do not consider values of $\alpha > 1$, for then even weak solutions of the Navier-Stokes
equations are not known to exist.
While the analysis in this section is not restricted to strictly fractal forces, the results apply
nevertheless to those with power-law Fourier coefficients $|\fk| \sim k^\zeta$ where
$\zeta = \alpha - \frac32 - \delta$ for any $\delta > 0$.

Write the steady body force as 
\begin{equation}
\vectf(\vectx) = F \vectphi(\ell^{-1} \vectx), 
\end{equation}
where $F$ is the amplitude of the force, the $H^{-\alpha}$ norm of~$\vectf$, 
and the ``shape'' function $\vectphi$ is a dimensionless 
divergence-free field on the unit 3-torus normalized
according to
\begin{equation}
\|\vectphi\|_{H^{-\alpha}} =1.
\end{equation}
Using Cauchy-Schwarz 
and the interpolation inequality \eqref{eq:interpol} 
with $s=\alpha$, $r=1$, $t=0$, we estimate
\[
\begin{split}
\left| \frac{1}{\ell^3} \int_{\ell^3} \vectf \cdot\vectu \, \rmd \vectx \right|
& \leq 
\sum_\vectk |\fk| \, |\uk| 
\leq 
\left[ \sum_\vectk (\ell k)^{-2\alpha} |\fk|^2\right]^{1/2} 
\left[ \sum_\vectk (\ell k)^{2\alpha} |\uk|^2\right]^{1/2} \\
& =
\|\vectf\|_{H^{-\alpha}} \|\vectu\|_{H^{\alpha}} 
= 
F \|\vectu\|_{H^{\alpha}} 
\leq 
F \|\vectu\|_{H^1}^\alpha \|\vectu\|^{1-\alpha}.
\end{split}
\]
Then taking time average
and applying H\"older's inequality,
\[
\begin{split}
\epsilon 
& \leq 
|\ave{\vectf\cdot\vectu}| 
\leq 
F \overline{\|\vectu\|_{H^1}^\alpha \|\vectu\|^{1-\alpha}} 
\leq 
F \left(\overline{\|\vectu\|_{H^1}^2}\right)^{\frac{\alpha}{2}} 
\left(\overline{\|\vectu\|^{\frac{2(1-\alpha)}{2-\alpha}}}
	\right)^{\frac{2-\alpha}{2}} \\
& =  
F \left(\frac{\nu}{\ell^2}\right)^{-\frac{\alpha}{2}} 
\left(\frac{\nu}{\ell^2}\overline{\|\vectu\|_{H^1}^2}
	\right)^{\frac{\alpha}{2}} 
\left(\overline{\|\vectu\|^{\frac{2(1-\alpha)}{2-\alpha}}}
	\right)^{\frac{2-\alpha}{2}} \\
& = 
F \left(\frac{\nu}{\ell^2}\right)^{-\frac{\alpha}{2}} 
\epsilon^{\frac{\alpha}{2}} 
\left(\overline{\|\vectu\|^{\frac{2(1-\alpha)}{2-\alpha}}}
	\right)^{\frac{2-\alpha}{2}} 
\ .
\end{split}
\]
Note that $\frac{(1-\alpha)}{2-\alpha}\in [0,\frac12]$  so Jensen's  
inequality (\ref{eq:J1}) ensures that the last term in the last line
above is bounded by $U^{1-\alpha}$.
Hence
\begin{equation}
\epsilon 
\leq 
\ell^{\frac{2\alpha}{2-\alpha}} \, 
\nu^{-\frac{\alpha}{2-\alpha}} \, 
F^{\frac{2}{2-\alpha}} \, 
U^{\frac{2(1-\alpha)}{2-\alpha}} \ .
\label{elphaba}
\end{equation}

On the other hand we can also estimate $F$ from above independently
in terms of $U$, $\nu$ and $\ell$.
Multiply the Navier-Stokes equation (\ref{NSE}) by a sufficiently smooth
time-independent, divergence-free function~$\vectpsi(\ell^{-1}\vectx)$  on the unit 3-torus 
satisfying $\ave{\vectphi \cdot \vectpsi} > 0$.
(It's easy to produce such fields $\vectpsi$, for example as a finite Fourier mode Galerkin 
truncation of $\vectphi$.)
Integrating by parts, taking time averages, and applying H\"older and Cauchy-Schwarz,
\begin{equation}  \label{eq:eps1}
F\ave{\vectphi \cdot \vectpsi}
= -\ave{\vectu \cdot (\nabla \vectpsi) \cdot \vectu} -  \nu \ave{\vectu\cdot\Delta \vectpsi}
\leq \|\nabla \vectpsi\|_{L^\infty} \overline{\|\vectu\|^2} +
\nu \|\Delta\vectpsi\| \overline{\|\vectu\|^2}^{1/2} .
\end{equation}
Hence
\begin{equation}
F \leq \frac{1}{\ave{\vectphi \cdot \vectpsi}} 
\left[
\|\tilde{\nabla} \vectpsi\|_{L^\infty} \frac{U^2}{\ell} 
+  \|\vectpsi\|_{H^2} \frac{\nu U}{\ell^2} 
\right] \ , 
\end{equation}
where $\tilde{\nabla} = \ell \nabla$ is the dimensionless gradient on the unit 3-torus .
Plugging this estimate for $F$ into the bound (\ref{elphaba}) for $\epsilon$ we deduce 
\begin{equation}  \label{eq:time-indep}
\dissipation 
\leq 
Re^{\frac{\alpha}{2-\alpha}} \left(
C_1 + C_2 \, Re^{-1}
\right)^{\frac{2}{2-\alpha}} \ ,
\end{equation}
where the coefficients $C_j$ depend only on the shape function $\vectphi$
and the multiplier function $\vectpsi$---but {\it not}
on the parameters of the problem, i.e., the force strength $F$, 
the viscosity $\nu$, or the outer length scale $\ell$.
Specifically,
\begin{equation}
C_1 = \frac{\|\vectphi\|_{H^{-\alpha}}^2 \| \tilde\nabla \vectpsi \|_{L^\infty}^2}
{\ave{\vectphi \cdot \vectpsi}^2} \quad \text {and} \quad
C_2 = \frac{\|\vectphi\|_{H^{-\alpha}}^2 \| \vectpsi \|_{H^2}^2}
{\ave{\vectphi \cdot \vectpsi}^2}.
\end{equation}

For $Re \gg 1$ the upper bound \eqref{eq:time-indep} scales
\begin{equation}
\dissipation \lesssim Re^{\frac{\alpha}{2-\alpha}}
\end{equation}
where the exponent $\frac{\alpha}{2-\alpha} \in [0,1]$.
If $\alpha =0$, i.e., when the force $\vectf \in L^2$, we recover
the classical estimate corresponding to Kolmogorov scaling
\begin{equation}
\dissipation \lesssim 1
\end{equation}
that holds as well when $\alpha < 0$  \cite{DoeringF02}.
In the other borderline case $\alpha=1$,
\begin{equation}
\dissipation \lesssim Re.
\end{equation}
And as advertised, when $Re \ll 1$ the overdamped laminar scaling
\begin{equation}
\dissipation \lesssim Re^{-1}
\end{equation}
emerges for all $\alpha \le 1$.  

\section{A time dependent fractal force}

Following Mazzi \& Vassilicos \cite{MazziV04}, consider a fractal forcing 
function of the form: 
\begin{equation}  \label{eq:Fk}
\fk(t) = F \ (\lk)^{\zeta -\delta} \left(\frac{\uk}{|\uk|} 
	+ \rmi \frac{\vectk\times\uk}{|\vectk|\,|\uk|} \right) \ ,
\end{equation}
where $F$ is the strength coefficient, $\zeta\in[-\frac32,-\frac12]$ and $\delta > 0$
and $\fk \equiv 0$ whenever $\uk=0$.
The Navier-Stokes equations (\ref{NSE}) driven by this velocity-dependent
time-varying force constitute an autonomous system.
We assume initial data $\vectu_0(\vectx) \ne 0$, that a 
(statistically) steady flow is subsequently sustained,
and that for $t>0$
each $|\uk(t)|=0$ only on a measure-zero set of times.
The scaling exponent $\zeta = -\frac12$ corresponds to the case 
where the forcing is in $H^{-1}$ at each instant of time
for all $\delta > 0$, while $\zeta = -\frac32$ (or less)
is $L^2$ (or smoother) forcing.  

Start by writing
\begin{align}
\epsilon  & =  \ave{\vectf \cdot \vectu} = \overline{\sumk \fk\cdot\uk^*}
\nonumber
\\ 
&= F \ \overline{ \sumk\frac{1}{(\lk)^{3/2+\delta}} 
	(\lk)^{\zeta+3/2}|\uk| }.
\end{align}
Applying the Cauchy-Schwarz inequality,
\begin{equation}  
\epsilon \leq F \, C \, \overline{\left(\sumk(\lk)^{2\zeta+3}|\uk|^2\right)}^{1/2} .
\end{equation}
where
\begin{equation}
C \equiv \left( \sumk\frac{1}{(\lk)^{3+2\delta}} \right)^{1/2}.
\label{C}
\end{equation}
Note that the ($3$-$d$) sum defining $C$ converges  iff $\delta > 0$.
Indeed, $C = {\cal O}(\delta^{-1/2}) \ \text{as} \ \delta \rightarrow 0$.
H\"older's inequality then implies
\begin{align}
\epsilon &\leq C F \left( \overline{ \sumk|\uk|^2}\right)^{-\zeta/2-1/4} \ 
	\left( \overline{\sumk (\lk)^2|\uk|^2} \right)^{\zeta/2+3/4} \\
	&= C F U^{-\zeta-1/2} \left(  \frac{\epsilon \ell^2}{\nu} \right)^{\zeta/2 + 3/4}.
\end{align}
Solving for $\epsilon$,
\begin{equation}
\epsilon \leq C^{\frac{4}{1-2\zeta}} \,
F^{\frac{4}{1-2\zeta}} \,
U^{\frac{4\zeta+2}{2\zeta-1}} \left(  \frac{\nu}{\ell^2} \right)^{\frac{2\zeta+3}{2\zeta-1}}.
\label{EFF}
\end{equation}
Now the challenge is to eliminate $F$ in favor of $U$, $\ell$ and $\nu$.

To derive an upper bound on $F$ we will estimate the bulk average of the
\eqref{NSE} dotted into the time-dependent test function $\vectpsi(\vectx,t)$ 
with the Fourier coefficients:
\begin{equation}
\psik(t) = \frac{\uk}{|\uk|} (\lk)^{-4-\delta'}
\end{equation}
for $|\uk| \ne 0$, with $\psik = 0$ when $|\uk| = 0$, and $\delta' > 0$.
We consider the resulting terms one by one.

First note that the pressure term $\ave{\vectpsi \cdot \nabla p }=0$ since $\nabla \cdot \vectpsi = 0$. 
The advection term is estimated
\begin{equation}
|\ave{\vectpsi \cdot (\vectu \cdot \nabla \vectu)}| 
= |\ave{\vectu \cdot (\nabla \vectpsi) \cdot \vectu}|
\leq \|\nabla \vectpsi\|_{L^{\infty}} \ave{|\vectu|^2}
\end{equation}
where 
\begin{equation}
\| \nabla \vectpsi \|_{L^{\infty}}  \le  \ell^{-1} \sumk \lk \, |\psik(t)|
\le \ell^{-1} \sumk (\lk)^{-3-\delta'} = \frac{C'}{\ell}
\end{equation}
and the pure number $C'$ is finite for all $\delta' > 0$; $C' = {\cal O}(\delta'^{-1})$ 
as $\delta' \rightarrow 0$.
The force term is
\begin{equation}
\ave{\vectf \cdot \vectpsi} = F\sumk (\lk)^{\zeta-4-\delta - \delta'} = C'' F
\label{boosk}
\end{equation}
where the sum for $C''$ converges uniformly for all non-negative $\delta$ and $\delta'$
and all $\zeta \le -1/2$.
By asserting {\it equality} in (\ref{boosk}) above we use the assumption
that $|\uk|>0$ for almost all $t>0$.
Next, the viscous term is
\begin{equation}
\ave{\vectpsi \cdot \nu \Delta \vectu } = 
\nu \ave{\Delta \vectpsi \cdot \vectu } \le
\nu \|\vectpsi\|_{H^2} \, U =
C''' \, \frac{\nu U}{\ell^2}
\end{equation}
where
\begin{equation}
C''' = \left( \sumk \lk^{-4-2\delta'} \right)^{1/2}
\end{equation}
is uniformly bounded for all $\delta' \ge 0$.
Finally, observe that
\begin{equation}
\ave{\vectpsi \cdot \partial_t \vectu } = \overline{\ddt \sumk |\uk|(\lk)^{-4-\delta'}}.
\end{equation}
The time average of the time derivative of a quantity vanishes if the quantity
is uniformly bounded in time.
Because
\begin{equation}
\sumk |\uk(t)|(\lk)^{-4-\delta} \le \left( \sumk |\uk(t)|^2 \right)^{1/2}
 \left( \sumk (\lk)^{-8-2\delta'}  \right)^{1/2}
\end{equation}
where $\sumk |\uk(t)|^2 = \|\vectu(\cdot,t)\|^2$ is uniformly bounded in time for these $H^{-1}$ (or smoother) forces, the sum above converges for all $\delta' \ge 0$ and
we conclude that $\ave{\vectpsi \cdot \partial_t \vectu } = 0$.

Hence the bulk average of $\vectpsi$ dotted into the Navier-Stokes equations yields
\begin{equation}
F \le \frac{C'}{C''} \, \frac{U^2}{\ell} +  \frac{C'''}{C''} \frac{\nu U}{\ell^2}
\end{equation}
with absolute constants $C''$ and $C'''$, and $C'$ depending only on $\delta' > 0$.
Inserting into (\ref{EFF}),
\begin{equation}
\dissipation \leq Re^{\frac{3+2\zeta}{1-2\zeta}} \,
\left( c_1 + c_2 Re^{-1} \right)^{\frac{4}{1-2\zeta}}
\end{equation}
where
\begin{equation}
c_1 = \frac{C C'}{C''} \quad \text{and} \quad
c_2 = \frac{C C'}{C''}.
\end{equation}

As before, when $Re \rightarrow 0$, this result produces the laminar scaling
\begin{equation}
\dissipation \lesssim Re^{-1},
\end{equation}
for all relevant values of the force's scaling exponent.
When $Re \rightarrow \infty$, however, the dissipation may be as large as
\begin{equation}
\dissipation \lesssim Re^{\frac{3+2\zeta}{1-2\zeta}}
\end{equation}
with exponent $0 \le \frac{3+2\zeta}{1-2\zeta} \le 1$ as $\zeta$
varies from $-\frac{3}{2}$ to $-\frac{1}{2}$.
It is worthwhile noting that the coefficients $c_1$ and $c_2$ depend on
$\delta > 0$ (and $\delta' > 0$, introduced for convenience)---but not at
all on the force parameters $F$ and $\zeta$ or on $\nu$ or $\ell$---and that the coefficients
$c_1(\delta)$ and $c_2(\delta)$ diverge as $\delta \rightarrow 0$
because $C(\delta)$ defined in (\ref{C}) diverges as $\delta \rightarrow 0$.


\section{Summary \& discussion}


In this paper we generalized the analysis that was previously employed for square integrable (or smoother) steady forces \cite{DoeringF02} and velocity-dependent forces  \cite{DoeringP05}
to derive bounds on the energy dissipation in the case of broad-band and
fractally-forced flow described by the incompressible
$3$-$d$ Navier-Stokes equations.
When a steady body-force $\vectf(\vectx) \in H^{-\alpha}$ with $\alpha \in [0,1]$, we showed
that the dimensionless dissipation factor $\dissipation(Re)$ is limited according to
\begin{equation}
4 \pi^2 Re^{-1} \le \dissipation 
\leq Re^{\frac{\alpha}{2-\alpha}} \left(
C_1 + C_2 Re^{\alpha-2}
\right)^{\frac{2}{2-\alpha}}.
\label{std}
\end{equation}
For velocity-dependent fractal forces of the form (\ref{eq:Fk}) 
with $|\fk| \sim k^{\zeta -\delta}$, $\zeta \in [-\frac32,-\frac12]$, and $\delta > 0$, we 
deduced that
\begin{equation}
4 \pi^2 Re^{-1} \le  \dissipation \leq Re^{\frac{3+2\zeta}{1-2\zeta}} \,
\left( c_1(\delta) + c_2(\delta) Re^{-1} \right)^{\frac{4}{1-2\zeta}}.
\label{vdf}
\end{equation}

These scalings are sharp as $Re \rightarrow 0$, displaying the laminar behavior 
$\dissipation \sim Re^{-1}$.
As $Re \rightarrow \infty$, both upper estimates are $\dissipation \sim Re^{0}$
for square integrable forcing, i.e.,  $\alpha = 0$ and $\zeta = -\frac32$.
And in the extreme limits $\alpha = 1$ and $\zeta = -\frac12$,
both estimates give $\dissipation \sim Re^{1}$.
We remark that the scalings in (\ref{std}) and (\ref{vdf}) are clearly consistent with each other
when it is recognized that forces with $|\fk| \sim k^{\zeta -\delta}$ are in 
$H^{-\alpha}$ when $\zeta = \alpha - \frac32$.

In terms of dimensional physical quantities, we have estimated the energy dissipation 
rate (per unit mass) $\epsilon$ in terms of the rms velocity $U$, $\ell$ and $\nu$.
Laminar dissipation corresponds to  $\epsilon \sim \nu U^2 / \ell^2$ while
the turbulent cascade is characterized by $\epsilon \sim U^3 / \ell$
and the roughest fractal forces may allows $\epsilon \sim U^4 / \nu$.
But for a specified form of the body force it is natural to consider
$\epsilon$ and $U$ as functions of the forcing amplitude $F$,
$\ell$ and $\nu$ \cite{ChildressKG01}.
When the force is specified, rather than the rms velocity, it is well known
(and easy to show) that the Stokes flow driven by the given 
force sets an upper limit for the dissipation rate; any other 
flow necessarily dissipates less energy.
In terms of the $F$, $\ell$ and $\nu$, the maximal Stokes flow dissipation is 
$\epsilon \sim  F^2 \ell^2 / \nu$ which may be interpreted as a laminar flow bound.
It is interesting to note that in the extreme limits of $H^{-1}$ forcing in 
(\ref{elphaba}) and $\zeta = -\frac12$ in (\ref{EFF}), the scaling
in this laminar upper limit is reproduced explicitly.


\begin{acknowledgments}
This research was funded in part by NSF grants PHY-0244859, DMS-0405903 and PHY-0555324.
CRD acknowledges additional support from a Humboldt Research Award, 
and NP is grateful for support from the Michigan Center for Theoretical Physics.
Part of this work was completed in the stimulating atmosphere of the Geophysical Fluid Dynamics 
Program at Woods Hole Oceanographic Institution.
\end{acknowledgments}


\newpage

\appendix

\section{Inequalities}

For convenience, in this appendix we collect the mathematical estimates used here:

\begin{itemize}
\item[(a)]
{\em Jensen's inequality}:  
If the real-valued function of a real variable $\theta(x)$  
is convex, then for each real-valued function $g$ 
\begin{equation}  \label{eq:Jensen}
\theta (\langle g \rangle ) 
\leq 
\langle \theta \circ g \rangle \ ,
\end{equation}
where $\langle \cdot \rangle$ stands for averaging.  
In particular, 
for any nonnegative function $g$ 
and any real number $p\in[0,1]$, 
\begin{equation}  \label{eq:J1}
\langle g^p \rangle
\leq 
\langle g \rangle^p
\ .
\end{equation}

\item[(b)]
{\em H\"older's inequality}: 
\begin{equation}  \label{eq:Holder}
\left|\int \vectphi(\vectx)\,\vectpsi(\vectx)\,\rmd\vectx\right| 
\leq 
\left(\int |\vectphi(\vectx)|^p\,\rmd\vectx\right)^{1/p}
\left(\int |\vectpsi(\vectx)|^q\,\rmd\vectx\right)^{1/q} 
\end{equation}
valid for all $\vectphi\in L^p$ and $\vectpsi\in L^q$, 
where $p$ and $q \ge 1$ and $\frac1p + \frac1q = 1$.  
For an $l^p$ sequence $(a_k)$ and an $l^q$ sequence 
$(b_k)$ (where $p$ and $q$ are related as above) 
the discrete analogue of \eqref{eq:Holder} reads
\begin{equation}  \label{eq:Holder-seq}
\left|\sum_k a_k b_k \right| 
\leq 
\left(\sum_k |a_k|^p \right)^{1/p}
\left(\sum_k |b_k|^q \right)^{1/q} \ .  
\end{equation}

An important case of \eqref{eq:Holder-seq}  
(for $p=q=2$) is the {\em Cauchy-Schwarz inequality}
\begin{equation}  \label{eq:two-ave}
\overline{\vectphi\vectpsi}
\leq 
\overline{\vectphi^2}^{1/2} 
\overline{\vectpsi^2}^{1/2} \ .
\end{equation}

\item[(c)]
{\em Interpolation inequalities} 
between Sobolev spaces: 
Let $0\leq r < s < t$ and $\vectu\in H^t$.
Note the algebraic identities
\[
r\,\frac{t-s}{t-r} + t\,\frac{s-r}{t-r} = s, 
\qquad 
\frac{t-s}{t-r} + \frac{s-r}{t-r} = 1.
\]
These interpolation estimates are the result of applying
H\"older's inequality~\eqref{eq:Holder-seq}
in Fourier space: 
\begin{eqnarray}  \label{eq:interpol}
\|\vectu\|_s^2 
&=& 
\sum_\vectk(\ell k)^{2s}|\uk|^2 
= 
\sum_\vectk \left[(\ell k)^{2r}|\uk|^2\right]^{\frac{t-s}{t-r}} 
\left[(\ell k)^{2t}|\uk|^2\right]^{\frac{s-r}{t-r}} 
\nonumber \\[2mm]
&\leq& 
\left[\sum_\vectk (\ell k)^{2r}|\uk|^2\right]^{\frac{t-s}{t-r}} 
\left[\sum_\vectk (\ell k)^{2t}|\uk|^2\right]^{\frac{s-r}{t-r}} 
\nonumber \\[2mm]
&=& 
\|\vectu\|_r^{2\frac{t-s}{t-r}}
\|\vectu\|_t^{2\frac{s-r}{t-r}} .
\end{eqnarray}

\end{itemize}


\bibliography{fractal_forcing}

\end{document}